# Multimode silicon photonics using on-chip geometrical-optics

Chunlei Sun,[†] Yunhong Ding,[§] Zhen Li,[†] Wei Qi,[†] Yu Yu,[†*] and Xinliang Zhang[†]

[†]Wuhan National Laboratory for Optoelectronics, Huazhong University of Science and Technology, Wuhan 430074, China

[§]Department of Photonics Engineering, Technical University of Denmark, Building 343, 2800 Kgs. Lyngby, Denmark

[*]Corresponding Author: yuyu@mail.hust.edu.cn

**ABSTRACT:**

On-chip optical interconnect has been widely accepted as a promising technology to realize future large-scale multiprocessors. Mode-division multiplexing (MDM) provides a new degree of freedom for optical interconnects to dramatically increase the link capacity. Present on-chip multimode devices are based on traditional wave-optics. Although large amount of computation and optimization are adopted to support more modes, mode-independent manipulation is still hard to be achieved due to severe mode dispersion. Here, we propose a universal solution to standardize the design of fundamental multimode building blocks, by introducing a geometrical-optics-like concept adopting waveguide width larger than the working wavelength. The proposed solution can tackle a group of modes at the same time with very simple processes, avoiding demultiplexing procedure and ensuring compact footprint. Compare to conventional schemes, it is scalable to larger mode channels without increasing the complexity and whole footprint. As a proof of concept, we demonstrate a set of multimode building blocks including crossing, bend, coupler and switches. Low losses of multimode waveguide crossing and bend are achieved, as well as ultra-low power consumption of the multimode switch is realized since it enables reconfigurable routing for a group of modes simultaneously. Our work promotes the multimode photonics research and makes the MDM technique more practical.

# 1. INTRODUCTION

Wavelength-division multiplexing (WDM) technique using many wavelengths and operating almost exclusively in the single-mode regime, is one of the most popular techniques in the past decades for both long-haul and short-reach optical interconnects. However, it supports limited bandwidth and is hard to meet the rapidly increasing bandwidth demand required in new services tackling massive data volumes. Space division multiplexing based on multi-core or few-mode fibers/waveguides has been a promising technique to avoid communication capacity crunch [1-4], as it enables each spatial channel to carry independent information in a shared wavelength link. In particular, high-density and monolithically integrated mode-division multiplexing (MDM) system using multimode waveguides is of great interest in terms of power consumption and link price.

Recently, many efforts have been involved to achieve great strides in the on-chip MDM community [5-26]. Among these, it is remarkable that the supported number of mode has reached to ten (dual polarizations) or eleven (single polarization), thanks to the effective investigations on mode multiplexer [12, 13]. Incommensurate with the significant development of mode multiplexer, other key multimode devices that enable bending, intersecting, splitting and switching have been the major bottleneck for a complete MDM system-on-chip. The typical waveguide width of conventional multimode devices is comparable with the working wavelength, and the wave feature of light in such waveguides is strong. Thus, it is difficult to design mode-independent devices since the performance varies greatly from mode to mode. Although a demultiplexing-processing-multiplexing procedure is straightforward and flexible, it behaves inefficiently with larger footprint [14-16], higher power consumption and complicated controls. Therefore, handling multiple modes simultaneously and efficiently on a chip is highly desirable. More recently, several solutions had been proposed to address the issue, while most of the reported multimode devices/circuits so far are restricted to manipulation of a small number (≤ 4) of mode [18-23]. In order to support more modes, the amount of computation and optimization increases exponentially. The reason behind is the inherent large mode dispersion. Moreover, various materials or fabrication processes are adopted for different multimode devices, and building blocks for MDM system have not been standardized [20, 24], making it obstructive to monolithic integration of large-scale MDM system.

Here, we introduce the geometrical-optics concept to design and standardize multimode building blocks. In this "on-chip-geometrical-optics" configuration with waveguide width larger than the wavelength, light propagates in the plane as in free-space, while still confined vertically in the waveguide. In this regard, the wave feature of light is weak, while the particle feature is dominant. Waveguide effect tends to be negligible in the plane, and thus various modes tend to be degenerate. Therefore, the problem of mode-independent implementation can be solved fundamentally, and light propagation can be considered as the ray trajectory in the media according to a set of geometrical rules, which had been employed to design and analyze bulky optical devices successfully.

Analogous to mirrors and cube beam splitters in free space optics, the on-chip waveguide bend and 3-dB coupler are proposed to achieve multimode routing using geometrical-optics-like concept. Additionally, the simplest scheme of crossing with two waveguides intersecting directly is also proposed, inspired by the phenomenon that two beams can cross without interacting with each other. On the basis of 3-dB coupler, a 2×2 multimode switch is obtained using a Mach-Zehnder interferometer (MZI) structure. The switch is really energy-efficient since it enables reconfigurable routing for a group of modes simultaneously without demultiplexing. In principle, these devices support as many modes as possible, provided that the waveguide width is enlarged to decrease the diffraction of high-order modes. Since the proposed solution avoids the conventional demultiplexing-processing-multiplexing procedure [14-17], the whole multimode components still have compact footprint. Although wide waveguides are adopted, the device footprint in our proposed scheme is comparable to or even more compact than conventional concept using single-mode waveguide array, especially for the complicated case processing multiple channels simultaneously. For instance, the proposed scheme possesses significant advantages in terms of footprint in the case of multimode switch. The concept gives a universal and general solution for mode scalability.

## 2. REALIZATION OF MULTIMODE DEVICES

### 2.1 MULTIMODE WAVEGUIDE CROSSING

The waveguide crossing is inevitable in optical networks to obtain a complex and advanced architecture[27, 28]. In free space, two intersecting beams do not interact with each other and continue to travel in the original direction. For the photonic integrated circuit with two waveguides placed orthogonally, however, the light will diverge severely in the intersection region and scatters into the bypass waveguide, resulting in high loss and crosstalk. Actually, the light suffers inherent pinhole diffraction in the waveguide, whose width equals to that of the pinhole. To date, most of the reported multimode waveguide crossings are based on self-image effect to decrease the propagation loss induced by pinhole diffraction. However, the mode scalability is a challenge. Optimization becomes more complicated and larger amount of computations is required.

Here, we adopt the scheme with two bus waveguides directly intersecting with each other to weaken pinhole diffraction, as shown in Figure 1a. Figure 1b shows the calculated diverging angle of different waveguide widths for four transverse electric (TE) modes at 1550 nm[29]. It can be seen clearly that the larger waveguide width is, the smaller diverging angle is. This agrees with the fact that wider waveguide has a larger pinhole and thus weaker diffraction. On the other hand, the higher-order mode has a larger diverging angle. For the conventional waveguide with width comparable to the working wavelength (i.e. 1-2 μm), the four modes have very large but different diverging angles. For the bus waveguide with a width much larger than the wavelength, the diverging angles of the four modes are very small and getting closer. In principle, the crossing can support countless modes as long as to widen the waveguide to decrease the diffraction of high-order modes. In this letter, the width of bus waveguide is comprehensively chosen as 10 μm for four modes to ensure weak waveguide effect and compact footprint. The four modes tend to be degenerate and behave similar features as in free space. For the demonstration convenience, we only give the results of the lowest- and highest-order modes. Figures 1c and 1d show the simulated electric field propagation distributions for $TE_0$ and $TE_3$, respectively. The two modes both stay unchanged at the through port, and almost invisible field is induced at the cross port.

To obtain MDM signals, the four-mode (de)multiplexer consists of three cascaded adiabatic couplers (ACs) to excite $TE_1$-$TE_3$ modes[30]. Reference MDM structure comprising only mode multiplexer and demultiplexer is fabricated on the same chip to fully characterize the performance of the crossing. By subtracting the loss of the grating couplers, the insertion losses for $TE_0$-$TE_3$ modes are lower than 1.5, 2, 2.5 and 3 dB, and the modal crosstalk is lower than -15, -12, -8 and -8 dB in the wavelength range of 1520-1600 nm. In order to maximize the measurement accuracy, test structures with ten crossings in cascade are used to extract the insertion loss. Figure 1e shows the microscope image of the fabricated multimode waveguide crossing. By subtracting the loss induced by the grating couplers, mode multiplexer and de-multiplexer, the normalized spectra of the cascaded crossings with W=10 μm are shown in Figure 1f. The average insertion losses of single crossing are about 0.03, 0.07, 0.16 and 0.75 dB for $TE_0$-$TE_3$ modes in the wavelength range of 1525-1600 nm, respectively. As a comparison, the cascaded crossings with W=6 μm are also fabricated and measured. The average insertion losses for $TE_0$-$TE_3$ modes are 0.05, 0.25, 0.31 and 1 dB, as shown in Figure 1g. The loss is much higher than that of 10 μm case, which agrees with the results of Figure 1a. To the best of our knowledge, this is the multimode crossing with the lowest loss being reported. Figures 1h and 1i illustrate the inter-modal crosstalk for $TE_0$ and $TE_3$ inputs with ten crossings in cascade. The crosstalk is lower than -12.5 dB.

## 2.2 MULTIMODE WAVEGUIDE BEND

Multimode waveguide bend is another building block of the MDM system to realize a compact footprint and flexible layout. In free space, a mirror is applied to turn a tight angle of transmitting light. Naturally, we can use the law of reflection to design the waveguide bend. Figure 2a exhibits the schematic of the proposed waveguide bend. Geometrical optics can be utilized to predict light trajectory accurately under short wavelength approximation. Due to the high refraction index contrast of silicon-on-isolator (SOI) waveguide, it is quite advantageous to form a mirror with the help of total internal reflection (TIR) effect. We calculated the critical angle θc of TIR for the four modes at 1550 nm, as shown in Figure 2b. For the bus waveguide with W=10 μm, the critical angles of TIR for the four modes tend to a uniform value of 30.5°. In order to decrease the bend loss, we choose an incident angle of 45° which is far beyond the critical angle. Figures 2c and 2d show the simulated electric field distribution in waveguide bends for $TE_0$ and $TE_3$ at 1550 nm, respectively. The two modes both stay almost unchanged at the output. Weak light scattering for $TE_3$ mode is mainly attributed to the pinhole diffraction, which can be alleviated by adopting a wider waveguide.

Test structures with twenty-four bends in cascade are used to decrease the measurement errors. The microscope image of the fabricated multimode waveguide bend are shown in the Figure 2e. The normalized spectra of the cascaded bends with W=10 μm and W=6 μm are shown in Figures 2f and 2g, respectively. For the 10 μm case, the average insertion losses of a single bend are 0.04, 0.12, 0.21 and 0.52 dB for $TE_0$-$TE_3$ modes in the wavelength range of 1525-1600 nm. For the 6 μm case, the average insertion losses are 0.06, 0.18, 0.73 and 1.46 dB. Again, the loss of 10 μm case is much lower than that of other reported bends to date, to the best of our knowledge. Figures 2h and 2i show the inter-modal crosstalk for $TE_0$ and $TE_3$ inputs with twenty-four bends in cascade. The crosstalk is lower than -12.5 dB.

## 2.3 MULTIMODE 3-DB COUPLER

3-dB coupler, which is widely used as a power splitter and combiner[31-33], is the building block in photonic integrated devices and circuits such as switches, modulators and 90° hybrids for coherent optical receivers. Conventional structures of optical 3-dB couplers are directional couplers, multiple mode interference (MMI) couplers and adiabatic couplers. The underlying operating principle of these structures is lateral evanescent coupling and mode interference. Very few investigations addressing multimode 3-dB coupler had been reported since the lateral evanescent coupling is very weak in multimode waveguides and the difference of modes coupling lengths are very significant.

In free space, cube beam splitter has been available commercially. It is composed of a pair of right angle prisms. A metallic-dielectric coating on the hypotenuse is used to cement the two prisms. It is straightforward to design an on-chip multimode 3-dB coupler using the same concept. Figure 3a exhibits the schematic of the proposed 3-dB coupler, including two TIR mirrors placed closely with each other. A trench is aligned 45° with respect to the waveguide, forming a frustrated TIR (FTIR) coupler [34, 35]. The trench width is set as G along the light propagation direction. The right angle prism corresponds to the TIR mirror, while the metallic-dielectric coating corresponds to the trench. At the first TIR interface, an evanescent field exponentially decays and extends into the trench perpendicularly. If the trench is narrow enough, the field amplitude at the second trench wall will be non-negligible. Thus, a portion of the power will pass beyond the trench and the rest reflected at the first interface will be frustrated.

To realize a 50:50 splitting, we calculated the transmission efficiency with different trench widths for $TE_0$ and $TE_3$ modes, as shown in Figure 3b. Light is injected into the input port A and monitored at the cross-port and through-port. In the legends, "C" stands for cross-port, while "T" stands for through-port. For $TE_0$ mode, the width of the 3-dB coupler is 87 nm, and the width for $TE_3$ mode is 91 nm. The widths for $TE_1$ and $TE_2$ modes are in the range of 87-91 nm. Note that the transmission efficiency of 3-dB coupling point for $TE_3$ mode is lower than that of $TE_0$ mode due to higher loss induced by the pinhole diffraction. Therefore, the final width is chosen as 89 nm as a trade-off. The two modes are both split into two parts equally, and the mode profiles remain the same at the output, as shown in Figures 3c and 3d. Figure 3e illustrates the microscope image of the fabricated multimode 3-dB coupler. We measured the transmission profile of 3-dB coupler when inputting different modes. The normalized spectra are shown in Figures 3f-3i for $TE_0$-$TE_3$, respectively. The power imbalance between the cross-port and through-port is about 2.2 dB, 2.7 dB, 2.5 dB, 2.3 dB, and the insertion loss is about 1.2 dB, 1.4 dB, 1.8 dB, 2.4 dB, for $TE_0$-$TE_3$ mode. The power imbalance mainly results from the fabrication error of the trench width. Note that the power of cross-port is always larger than that of through-port, indicating the fabricated trench is wider than the designed one. One possible and effective way to improve the fabrication insensitivity is increasing the trench width by adopting the material with larger refraction index filled in the trench.

## 2.4 MULTIMODE 2×2 SWITCH

Based on the mode-transparent 3-dB couplers, it is quite easy to construct a 2×2 MZI-architecture multimode switch, as shown in Figure 4a. It consists of two 3-dB couplers connected by two same L-shape arms and four 90° bends. The switch performs two states, through and cross, which can be tuned by a π phase-shift in one arm of MZI via the index change induced by the thermal-optic effect. The metal Au electrode is utilized to connect the external electrical source with TiN heaters. Obviously, the switch is easy to be scaled in the plane to realize a compact and high-density integration due to its rectangular layout. To investigate the mode-transparence of the multimode switch, we calculate the phase-shift with different temperature variations for the four modes, as shown in Figure 4b. The length of MZI arm covered by the TiN heater is 160 μm, and the width of waveguide is 10 μm. It can be seen that the variation of phase-shift is indistinguishable from mode to mode under the same temperature. Therefore, the switch can allow dynamic routing of light transmission paths for a group of modes under a same drive power.

    Figure 4e shows the microscope image of the fabricated multimode switch, and the normalized response of the multimode switch are shown Figures 4d-4k. For demonstration convenience, the spectra are only given when light is injected into port A, and same results can be also obtained from port B. The legend "A0-D0-C" refers to the $TE_0$ mode transmission from port A to D in the cross-state, and the number "0" stands for the mode order. During the measurement of the switch, the driving voltages are set at a constant value for all modes. For $TE_0$

mode, the average insertion loss of port D is 0.9 dB in the through-state, as the black curve shown in Figure 4d. The average insertion loss is 6.5 dB when switching to cross-state, as the red curve shows. The other curves lower than -12.5 dB refer to the inter-modal crosstalk in both through-state and cross-state. Thus, the extinction ratio for through-cross switchover from port A to D is about 5.6 dB. As for the case of other modes ($TE_1$-$TE_3$) transmitted from port A to D, the switchover extinction ratios are about 6.2, 6.1 and 5.7 for modes, respectively. The results are shown in Figures 4e-4g. As for the transmission from port A to E, the switchover extinction ratios are approximately 15.1, 17.5, 15.2 and 17.4 dB for $TE_0$-$TE_3$ modes, and the results are shown in Figures 4h-4k, respectively. The device performs an obvious switch functionality, and the performance uniform for the four modes is really good. The switchover difference between the port D and port E is large, and it mainly results from the power imbalance of the 3-dB multimode coupler. We also measure the power consumption, which is only 35 mW.

**CONCLUSION**

To sum up, in analogy with the bulky optical devices working in free space, we propose and demonstrate a set of on-chip multimode building blocks such as waveguide crossing, bend and 3-dB coupler, adopting geometrical-optics-like solutions based on bus waveguides. To our best knowledge, the demonstrated multimode waveguide crossing and bend both have the simplest design and the lowest loss. It is also the first time to demonstrate a multimode switch based on a mode transparent 3-dB coupler that performs reconfigurable routing for $TE_0$-$TE_3$ modes at the same time with ultra-low power consumption. The general solution is so concise that we can easily upgrade the structures to tackle more modes by only adopting wider waveguides. Although bus waveguides are used in the MDM system, it still has a compact footprint since all the proposed components are mode-transparent, inherently avoiding the conventional demultiplexing, processing and re-multiplexing procedure. The work will open up the possibility of high-density integration in MDM for ultra-high bandwidth communications.

**MATERIALS AND METHODS**

**Numerical simulation method:** The electric field propagation distribution and transmission response of the multimode devices are calculated by using the 3D finite difference time domain (3D FDTD) numerical method. Scattering bound condition is taken into account and perfectly matched layer (PML) is used to surround the simulation domain.

**Device fabrication**: The proposed MDM devices were fabricated on an SOI platform with a 220-nm-thick top silicon layer and a 2-μm SiO2 buried oxide layer. The waveguide structures were formed by the electron beam lithography (EBL) and inductively coupled plasma (ICP) etching. For the multimode switch, the TiN heater is deposited on the top of one arm of the MZI to induce the phase difference. A 1-μm thick SiO2 cladding by plasma-enhanced chemical vapor deposition (PECVD) covers the entire device, forming a buffer layer between the heater and the waveguides. The metal Au electrode is sputtered on a separate layer, realizing the connection to the external electrical source with TiN heaters.

**Experimental Method:** In order to obtain the actual response of the device, a silicon waveguide directly connected by a pair of grating couplers (GCs) is also fabricated as a reference. A broadband light source around 1550 nm is launched into a polarization beam splitter to obtain the linearly polarized light. Being assisted by the polarization controller (PC), the maximum coupling efficiency of the GC can be achieved. A power meter and an optical spectrum analyzer (OSA) are utilized to monitor the output light.

**ASSOCIATED CONTENT**

**AUTHOR INFORMATION**

Corresponding Author

*Email: yuyu@mail.hust.edu.cn

**Author Contributions**

C.L.S. and Y.Y. conceived the idea. Y.H.D fabricated the devices. C.L.S., Z.L. and W.Q. designed and carried out the experiments, and performed the data analysis. C.L.S., Y.Y. and X.L.Z discussed the results. Y.Y. and X.L.Z. supervised the project. All authors reviewed the manuscript.

**Notes**

The authors declare no competing financial interest.

**ACKNOWLEDGMENTS**

This work was supported by the National Natural Science Foundation of China (Grant No. 61775073 and 61911530161), the Program for HUST Academic Frontier Youth Team (2018QYTD08).

**Figure 1.** Multimode waveguide crossing. (a) Schematic of waveguide crossing which consists of two bus waveguides directly intersecting with each other. (b) Calculated diverging angle versus waveguide width for $TE_0$-$TE_3$ modes. Simulated electric field propagation distribution of waveguide crossings for (c) $TE_0$ and (d) $TE_3$ modes at 1550 nm. (e) Microscope image of the fabricated multimode crossing. Measured spectra of multimode crossings with (f) W=10 μm and (g) W=6 μm in the wavelength range of 1525-1600 nm. The inset shows ten crossings in cascade. The spectra have been normalized by subtracting the loss induced by the grating couplers, mode multiplexer and mode de-multiplexer. Inter-modal crosstalk when (h) $TE_0$ and (i) $TE_3$ modes are injected into the cascaded ten crossings.

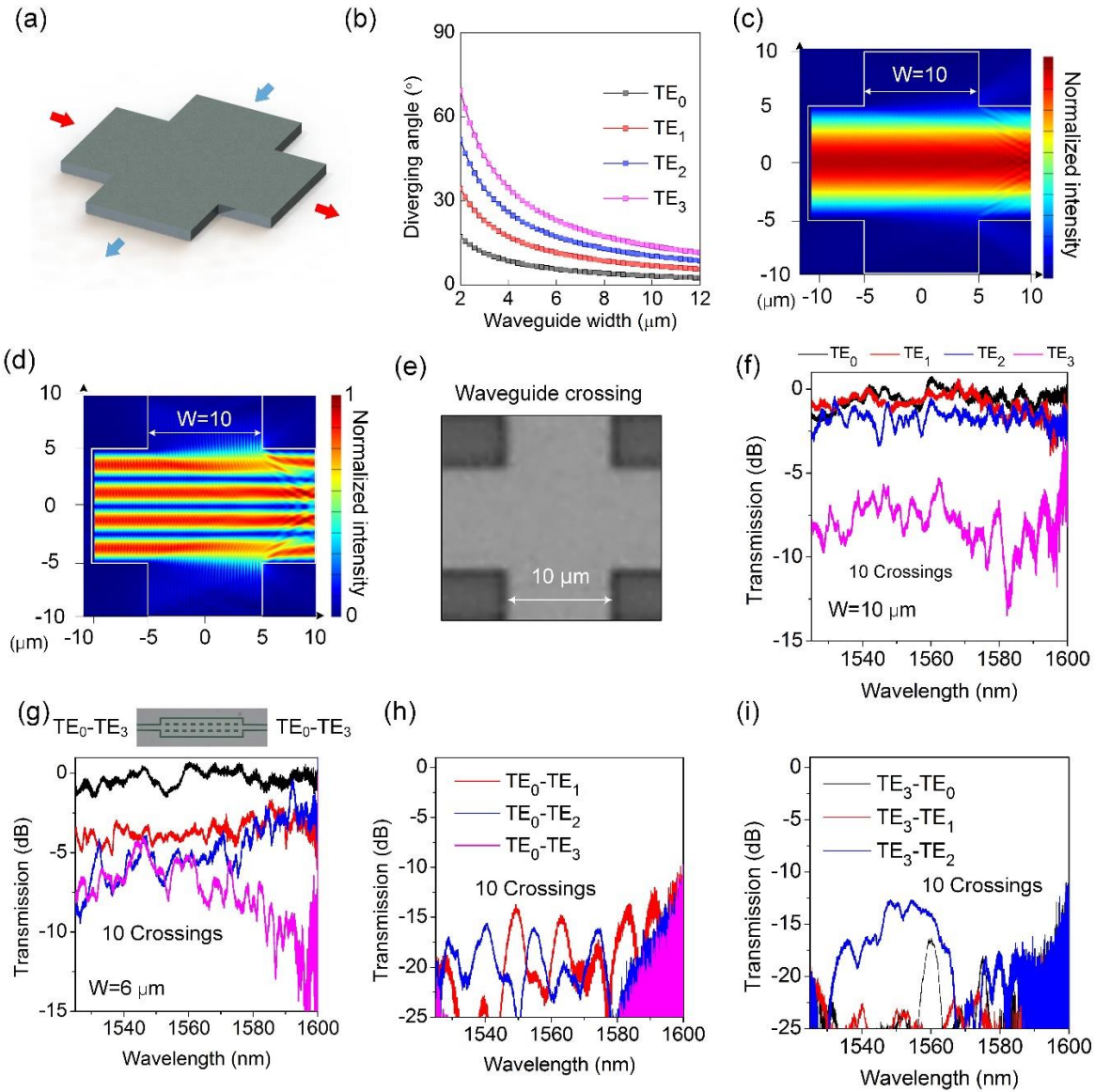

**Figure 2.** Multimode waveguide bend. (a) Schematic of waveguide bend with a TIR mirror aligned 45° with respect to the waveguide. (b) Calculated critical angle versus waveguide width for $TE_0$-$TE_3$ modes. Simulated electric field propagation distribution of waveguide bend for (c) $TE_0$ and (d) $TE_3$ modes at 1550 nm. (e) Microscope image of the fabricated multimode bend. Measured spectra of the multimode bend with (f) W=10 μm and (g) W=6 μm in the wavelength range of 1525-1600 nm. The inset shows a total of twenty-four bends in cascade. The inter-modal crosstalk when (h) $TE_0$ and (i) $TE_3$ modes are injected into the cascaded twenty-four bends.

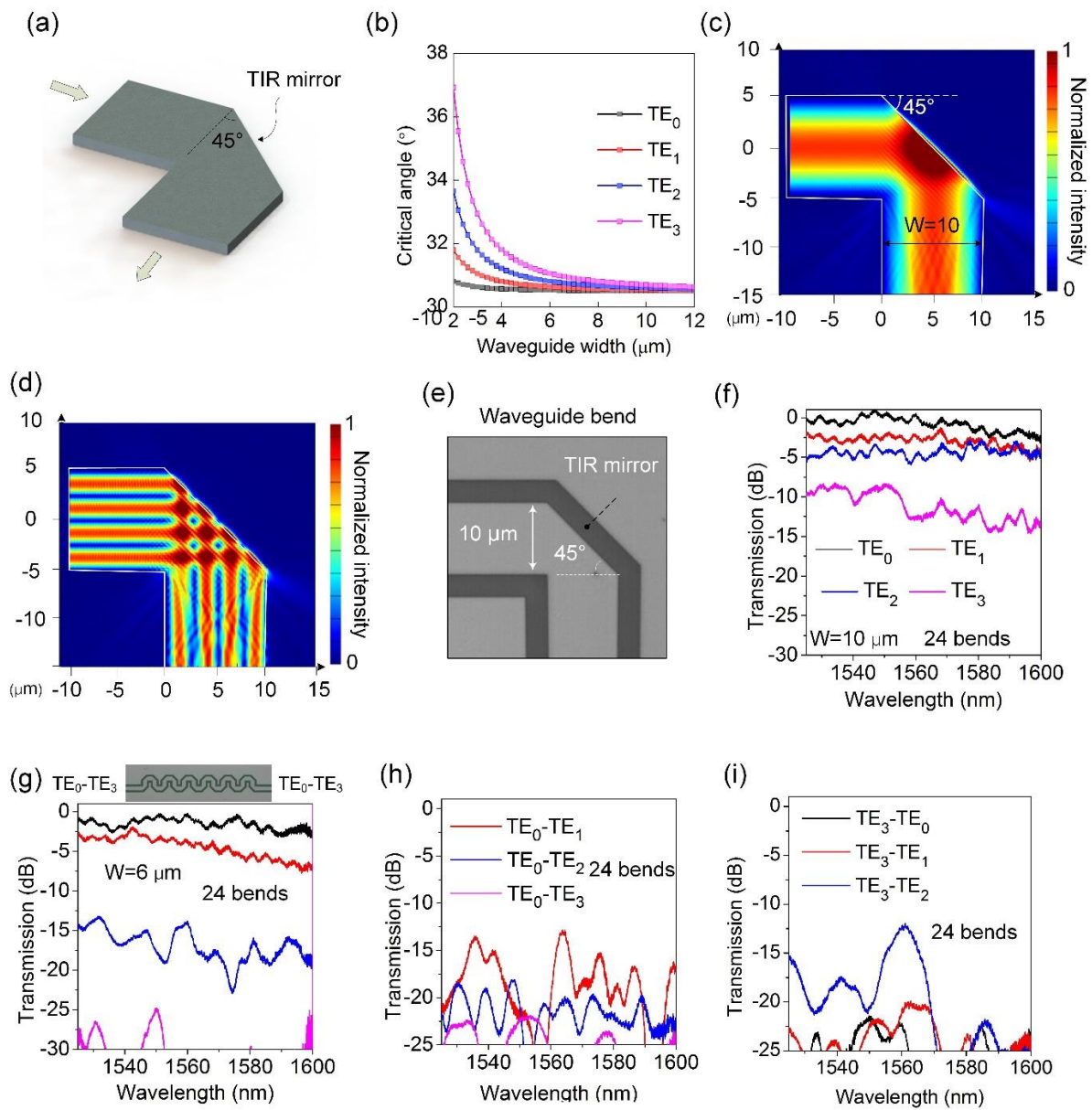

**Figure 3.** Multimode 3-dB coupler. (a) Schematic of 3-dB coupler which consists of two TIR mirrors separated by a narrow trench. (b) Simulated transmission efficiency with different trench widths for $TE_0$ and $TE_3$ modes at 1550 nm. The arrows indicate the 3-dB coupling points. Simulated electric field propagation distribution of waveguide bend for (c) $TE_0$ and (d) $TE_3$ modes at 1550 nm. The width of bus waveguide is 10 μm, and the trench width is 89 nm. (e) Microscope image of 3-dB multimode coupler. (f)-(i) Measured spectra of 3-dB coupler for $TE_0$-$TE_3$ modes at 1550 nm. In the legends, "C" refers to cross-port, while "T" refers to through-port.

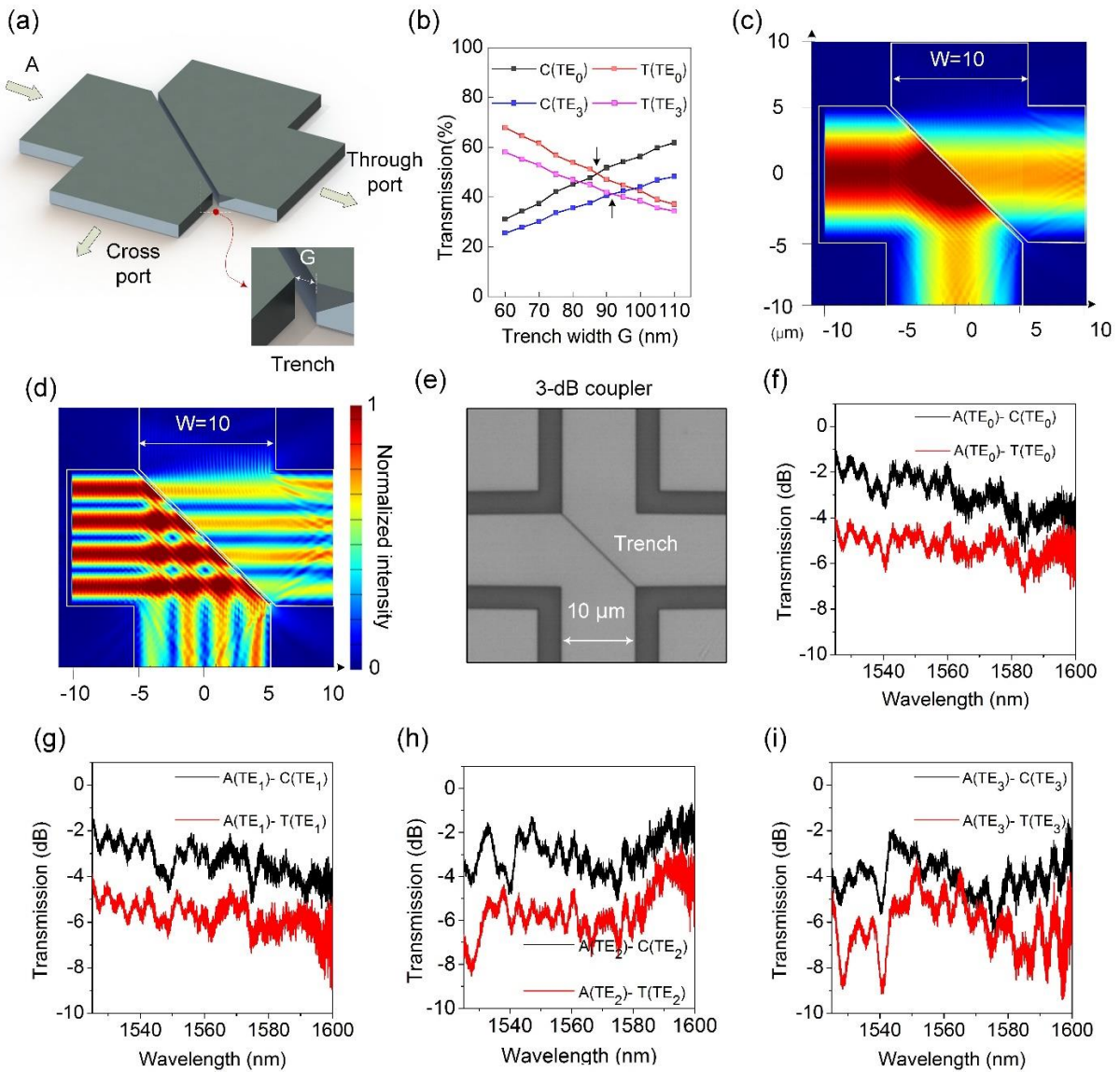

**Figure 4.** Multimode switch. (a) Schematic of the multimode switch which consists of two 3-dB couplers connected by two same L-shape arms and four waveguide bends. (b) Calculated phase-shift with different temperature variations for $TE_0$-$TE_3$ modes. (c) Microscope image of multimode switch. (d)-(k) normalized response of the multimode switch. The spectra are obtained when light is injected into port A and monitored at port D and port E. The switch works in two states: cross and through. The legend "A0-D0-C" stands for the $TE_0$ mode transmission from port A to D at the cross-state.

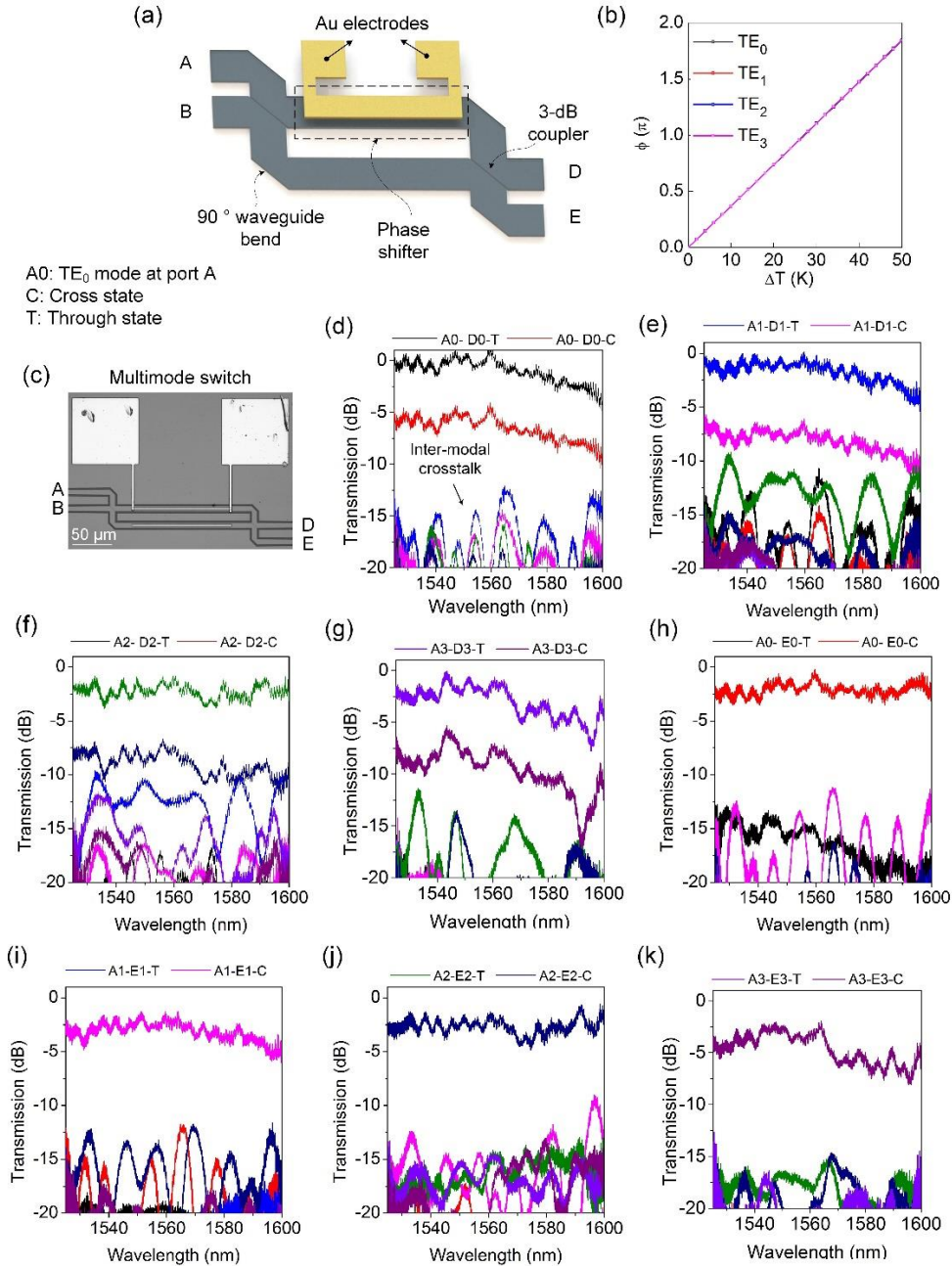

**For Table of Contents Use Only**
TOC Graphic

# Multimode silicon photonics using on-chip geometrical-optics

Chunlei Sun,[†] Yunhong Ding,[§] Zhen Li,[†] Wei Qi,[†] Yu Yu,[†*] and Xinliang Zhang[†]

A universal solution to standardize the design of fundamental multimode building blocks, by introducing a geometrical-optics-like concept. This work promotes the multimode photonics research and makes the MDM technique more practical.

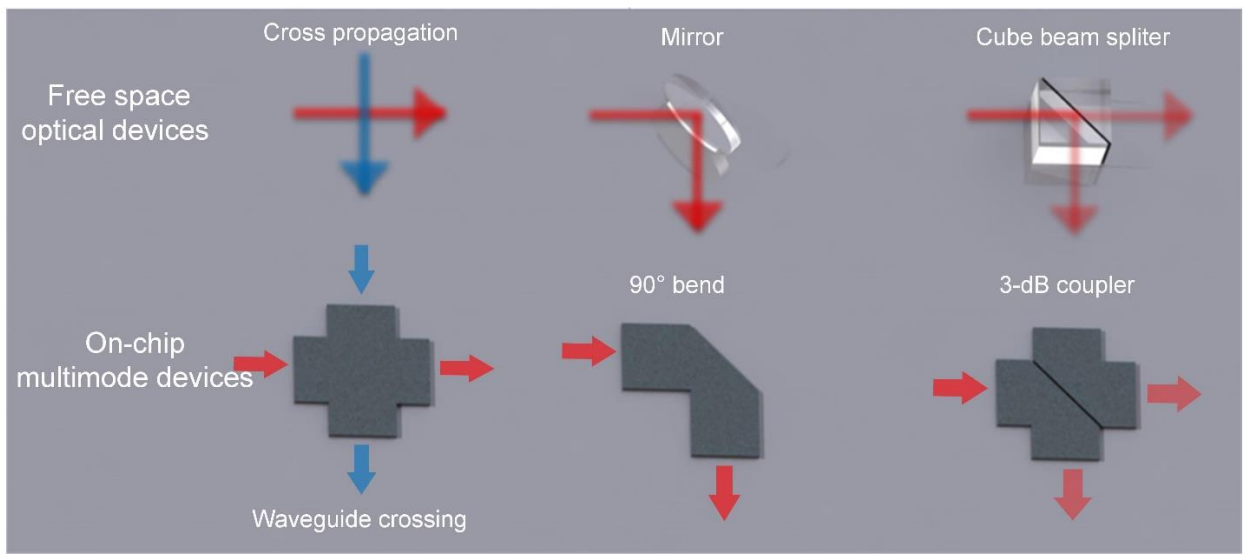